\documentclass{aastex631}

\usepackage{amsmath}
\usepackage{color}

\begin{document}

\title{Topological approach to void finding applied to the SDSS galaxy map}

\author{Manu Aggarwal}
\affiliation{Laboratory of Biological Modeling/NIDDK, National Institutes of Health, Bethesda, Maryland, United States of America}
\email{manu.aggarwal@nih.gov}
\author{Motonari Tonegawa}
\affiliation{Asia Pacific Center for Theoretical Physics, Pohang, 37673, Korea}
\email{motonari.tonegawa@apctp.org}
\author{Stephen Appleby}
\affiliation{Asia Pacific Center for Theoretical Physics, Pohang, 37673, Korea}
\affiliation{Department of Physics, POSTECH, Pohang, 37673, Korea}
\email{stephen.appleby@apctp.org}
\author{Changbom Park}
\affiliation{School of Physics, Korea Institute for Advanced Study, 85
Hoegiro, Dongdaemun-gu, Seoul, 02455, Korea}
\email{cbp@kias.re.kr}
\author{Vipul Periwal}
\affiliation{Laboratory of Biological Modeling/NIDDK, National Institutes of Health, Bethesda, Maryland, United States of America}
\email{vipulp@niddk.nih.gov}

\begin{abstract}
    The structure of the low redshift Universe is dominated by a multi-scale void distribution delineated by filaments and walls of galaxies. The characteristics of voids; such as morphology, average density profile, and correlation function, can be used as cosmological probes. However, their physical properties are difficult to infer due to shot noise and the general lack of tracer particles used to define them. In this work, we construct a robust, topology-based void finding algorithm that utilizes Persistent Homology (PH) to detect persistent features in the data. We apply this approach to a volume limited sub-sample of galaxies in the SDSS I/II Main Galaxy catalog with the $r$-band absolute magnitude brighter than $M_r=-20.19$,
    and a set of mock catalogs constructed using the Horizon Run 4 cosmological $N$-body simulation.
    We measure the size distribution of voids, their averaged radial profile, sphericity, and the centroid nearest neighbor separation, using conservative values for the threshold and persistence. We find $32$ topologically robust voids in the SDSS data over the redshift range $0.02 \leq z \leq 0.116$, with effective radii in the range $21 - 56 \, h^{-1} \, {\rm Mpc}$. The median nearest neighbor void separation is found to be $\sim 57 \, h^{-1} \, {\rm Mpc}$, and the median radial void profile is consistent with the expected shape from the mock data.
\end{abstract}

\section{Introduction} \label{sec:intro}

The cosmic web, as traced by galaxies in the late Universe, is an interesting feature of the
gravitational collapse of an initially Gaussian density field that naturally appears and evolves in the late-time universe \citep{1978ApJ...222..784G,10.1093/mnras/185.2.357,10.1093/mnras/204.3.891,1987ApJ...313..505W,1986ApJ...306..341G,1994ApJ...420..525V,Bond:1995yt,2005MNRAS.359..272C,Weygaert2008,10.1093/mnras/sty1643,Libeskind:2017tun,park2022formation}. In three dimensions, matter collapses
locally in an anisotropic manner, from underdensities onto two-dimensional sheets, then into
one-dimensional filaments and finally accreting into knots. Gravitational outfall from underdense
regions generates large voids in the late Universe, bounded by the two-dimensional walls generated
as a result of anisotropic collapse. Voids comprise the overwhelming majority of the spatial volume
at $z \simeq 0$ \citep{pan2012cosmic}. 

The defining characteristic of a void is a dearth of matter, which makes it difficult to quantify
and measure. We observe galaxies, which comprise a relatively sparse point distribution that is biased relative to the dark matter field.
Regions of space which are not well sampled will be heavily affected by noise when attempting to
infer properties of the underlying matter density field. In spite of these difficulties, voids have
proved to be a valuable source of cosmological information, and they have been applied to the
Alcock-Paczynski test \citep{Ryden:1995,lavaux2012precision, Sutter:2012tf, Sutter:2014oca}, and other forms of cosmological parameter estimation \citep{Hamaus:2015yza, Hamaus:2020cbu, Schuster:2022ogh, Contarini:2022mtu, Contarini:2022nvd,10.1093/mnras/stac2011}. Voids can also be used to test extensions to the standard model \citep{Lee_2009,Platen:2007ng,Bos:2012wq, PhysRevLett.111.241103, 10.1093/mnras/stv777, PhysRevD.94.103524, Woodfinden:2022bhx, DES:2018cnw, Verza:2019tvg, Massara:2015msa, PhysRevD.99.121304,10.1093/mnras/staa3074}.

To extract cosmological information from the distribution of voids, we must first measure their individual properties such as volume and morphology. Numerous different void finders are employed within cosmology for this purpose \citep{Vogeley1994, el1997voids, Platen:2007qk,neyrinck2008zobov,Aragon-Calvo:2010igh, pan2012cosmic,SUTTER20151}. They typically involve searching for
underdense regions with a fixed shape template (spherical, ellipsoidal), or generating a set of
polygons from galaxy positions, assigning a local density to these shapes according to their volume and finally linking adjacent, low-density polygons to generate macroscopic structures (but see \cite{Shim:2020wyj} and \cite{Shim:2023zgf} for a void definition not assuming any geometry). They yield somewhat different morphologies, density profiles etc, leading to some ambiguity in void properties. These different approaches are typically based on geometric assumptions and are fundamentally different from the mathematically rigorous topology computed by Persistent Homology (PH). The method of PH applies techniques developed in
algebraic topology to find robust lacunae in noisy, discrete data sets.  PH makes no assumptions about the geometry of voids \textit{a priori}, which is important since the real universe consists of regions of relatively low density that are more often polyhedral than spherical~\citep{neyrinck2008zobov,icke1987fragmenting}. Indeed, we do not expect voids to be spherical because the critical points of the Gaussian initial density field are generically ellipsoidal \citep{1986ApJ...304...15B}. As underdense regions grow in volume due to gravitational collapse onto surrounding walls, they have a tendency to become increasingly spherical. However, at late times they merge to form complex morphological structures \citep{Sheth:2003py,10.1093/mnras/stt1169}. 

In this work, we apply the PH methodology to the SDSS main galaxy sample, inferring a number of key
properties of voids in the low redshift universe; their size distribution, averaged radial profiles,
sphericity, and the pairwise distribution of void voxels. We perform a comparative study between the
data and mock galaxy catalogs and between our approach and another void finder in the literature. 

The paper proceeds as follows. In Section \ref{sec:top} we provide a brief and non-technical review of some of the important aspects of the topological methodology employed in this work. In Section \ref{sec:data} we introduce the galaxy catalog from which voids are extracted and the mock data that is used for comparative purposes. Section \ref{sec:results} contains the main results of our analysis; the properties of voids found using PH in the data and mock catalogs. We compare our results with other void finders in the literature in Section \ref{sec:comparison}, and discuss our findings in Section \ref{sec:disc}. The appendix contains some of the more technical details of the numerical algorithms used in the main body of the paper. 


\section{Introduction to Persistent Homology}
\label{sec:top}

We begin with a brief review of some of the important underlying ideas used in topological data analysis, in particular for the void finding algorithm applied to galaxy data in Section \ref{sec:results}. The discussion is intended to be non-technical; further details can be found in the appendices.

\begin{figure}[tbhp!] \centering \includegraphics[width=\textwidth]{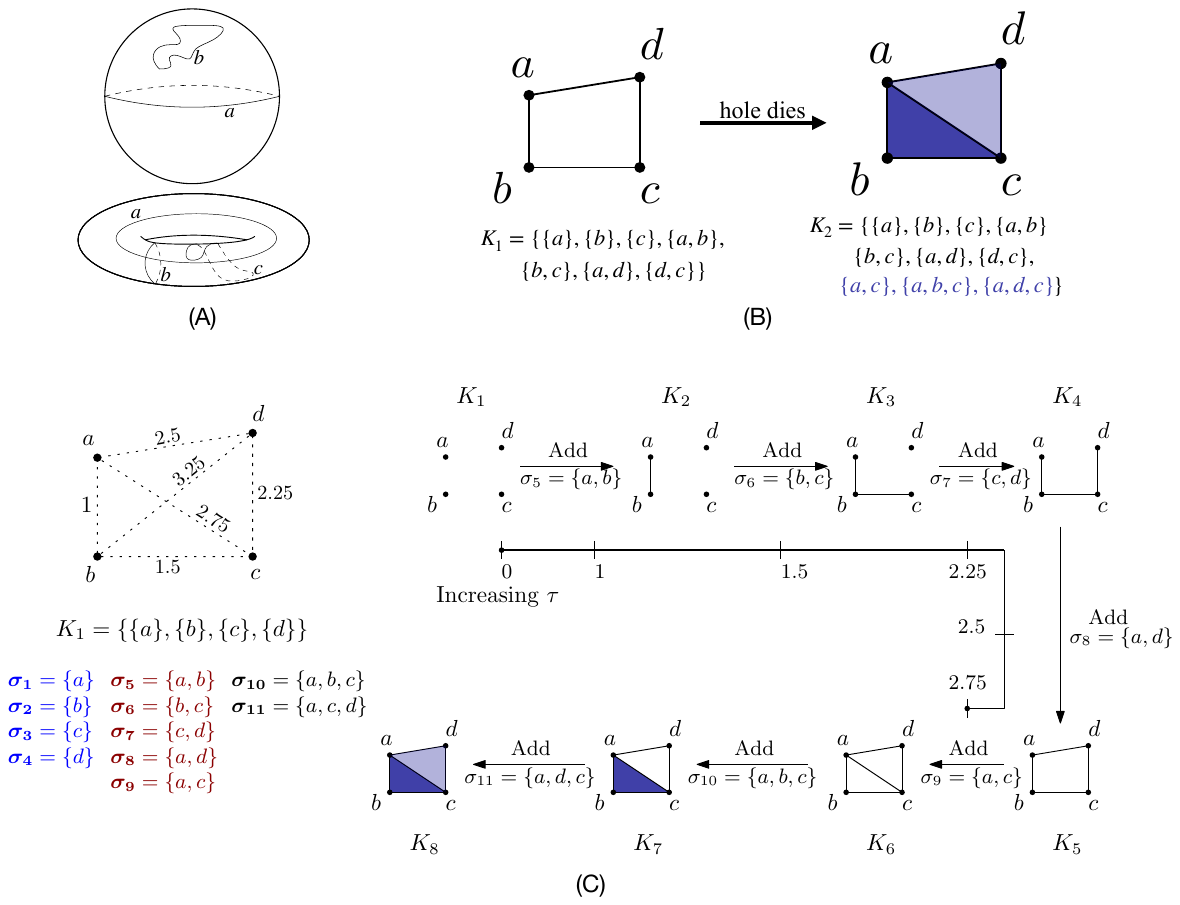}

  \caption{\label{fig:intro0}(A) Top panel shows two cycles $a$ and $b$ on the surface of a sphere. Both can be
  continuously deformed along the surface of the sphere to the same point. Any cycle on the surface of a
  sphere can contract to a point. Bottom panel shows three cycles on the surface of a $2$-torus. Cycle $c$ contracts to
  a point. Cycles $a$ and $b$ are non-contractible and cannot be deformed to each other. Any
  non-contractible cycle on the surface of the torus can deform to either $a$ or to $b.$ (B) A hole
  dies when its triangulation gets filled-in. The hole in simplicial complex $K_1$ dies when edge
  $\{a,c\}$ and triangles $\{a,b,c\}$ and $\{a,d,c\}$ are added to it to give new complex $K_2.$ (C) (from \cite{aggarwal2023tight})
  Vietoris-Rips filtration on a discrete set of points. The simplicial complex at a given value of
  $\tau$ is a collection of all simplices with diameter at most $\tau.$ Number of holes is initially
  $0.$ One hole is born at spatial scale of $\tau=2.5$ when edge $\{a,d\}$ is added. Another is born
  when $\{a,c\}$ is added at $\tau=2.75.$ Both holes get filled in, or die, at $\tau=2.75$ when
  triangles ($2$-simplices) are added to the simplicial complex. The first hole has persistence
  $2.75-2.5 = 0.25.$ The second hole dies at the same spatial scale it is born and so has persistence 0.}

\end{figure}

A homology group is a
collection of sets of cycles such that any two cycles in the same set can be continuously deformed
into one another and those in different sets cannot. For example, all cycles on the surface of a
sphere can be shrunk or contracted continuously along its surface to a point (see Figure
\ref{fig:intro0}A top panel). On the other hand, Figure \ref{fig:intro0} also shows three homologically distinct
cycles on the surface of a $2$-torus (Figure \ref{fig:intro0}A bottom panel). Cycle $c$ can contract to a point, whereas cycles $a$ and $b$
can neither contract to a point nor can be deformed to each other.  We say that cycle $c$ is
contractible (belonging to the trivial homology), and $a$ and $b$ are non-contractible cycles in
different equivalence classes of the homology group of dimension one (H$_1$). Moreover, any
non-contractible cycle on the surface of a torus can be deformed either to $a$ or to $b.$ Hence,
there are exactly two topologically distinct holes in this shape, and $a$ and $b$ are examples of
representative boundaries for these. This gives a classification of the shape of the surface of a
$2$-torus based on the number of homologically distinct non-contractible cycles on its surface. In
this example, we discussed cycles on the surface of the torus which are also called cycles of
dimension one and belong to the homology group of dimension one. Similarly, the homology group of
dimension two (H$_2$) is the collection of sets of non-contractible cycles of dimension two. 
Intuitively, they can be thought of as non-contractible surfaces around voids in a point-cloud (i.e., a discrete set of points)
embedded in a three-dimensional Euclidean space.

In real-life applications, experimental data are often discrete observations that can be embedded as
a point-cloud and not a smooth manifold. Homology groups of a point-cloud can be computed by constructing simplices. Briefly, a $n$-simplex is a set of $n+1$ points. For example, 0-simplices are points, 1-simplices are edges, 2-simplices are triangles, and 3-simplices are tetrahedrons. A collection of simplices is called a simplicial complex. Figure \ref{fig:intro0}B shows two simplicial complexes. Complex $K_1$ has a hole of dimension $1$ and the edges in the simplex form a representative boundary around this hole. The complex $K_2$ contains triangles $\{a,b,c\}$ and $\{a,c,d\}$ along with the simplices of $K_1.$ Visually, these are shown as the two filled-in triangles in $K_2$ in the figure. Naively, we can note that the hole in $K_1$ is filled in when its triangulation is added to the complex. We say that the hole in $K_1$ dies when the triangles  $\{a,b,c\}$ and $\{a,c,d\}$ are added to it. Simplicial homology rigorously defines this notion of holes and their birth and death using algebraic topology, and it generalizes to high dimensions. Readers interested in the mathematical and computational details are directed to \cite{edelsbrunner2022computational}.

When a discrete set is embedded in a Euclidean space, we can compute its homology groups at different spatial scales. Vietoris-Rips filtration (VR-filtration) is a commonly used construction of simplicial complexes at different spatial scales. The simplicial complex at spatial scale $\tau$ is defined as the collection of all simplices that have $\tau$ as the maximal pairwise distance between their points.
As the
spatial scale changes, the simplicial complex changes, and there might be birth and death of
holes (Figure \ref{fig:intro0}C). PH computes these births
and deaths, and they are plotted as persistence diagrams
(PD).

They give a global multi-scale overview of the topology of the shape of the discrete data set. Persistence of a topological feature, or a hole, is defined as the duration 
$\tau$ between its death and birth. Features with a large persistence are usually deemed as topologically significant.
Due to its generality and computation of robust topological features in noisy data sets across multiple scales, PH has
found useful applications in areas as diverse as neuroscience~\citep{bendich2016persistent},
computational biology~\citep{mcguirl2020topological}, natural language
processing~\citep{zhu2013persistent}, the spread of contagions~\citep{taylor2015topological},
cancer diagnosis~\citep{nicolau2011topology, lawson2019persistent}, material
science~\citep{kramar2013persistence}, computer graphics~\citep{2018arXiv181112543B}, cosmology \citep{MELOTT19901,Mecke:1994ax,Kerscher:1998gs,10.1111/j.1365-2966.2011.18395.x, park2013betti, Pranav:2018pnu, Feldbrugge:2019tal, Wilding:2020oza, Pranav:2021ozq} among many
others. In the context of cosmology, homology groups are associated with different cosmic
environment types as follows. Connected components ($0$-dimensional homology groups, H$_0$), loops
($1$-dimensional homology groups, H$_1$), and low-density 3D volumes ($2$-dimensional homology
groups, H$_2$) are analogous to galaxy clusters, closed loops of filaments, and cosmic voids,
respectively~\citep{xu2019finding}. Locations of voids can be estimated by computing representative
boundaries of topologically significant H$_2$ features. However, the representatives are not unique
by definition~\citep{carlsson2009topology}. We define tight representatives as those with shorter
lengths (fewer number of simplices in the boundary). Computing optimal tight representatives is computationally intractable even for a few thousand points.~\cite{aggarwal2023tight} developed an algorithm for large data sets that computes representatives that may not be optimal but are significantly shortened as compared to those obtained from the matrix reduction algorithm by default. Using it, we find geometrically precise boundaries around significant topological features in the SDSS and HR4 data sets which have more than a hundred thousand galaxies.

\section{Data} \label{sec:data}

\subsection{SDSS spectroscopic sample}\label{sec:data_SDSS}

To construct void catalogs we require a galaxy catalog, preferably with certain desirable
properties. The data should be contiguous on the sky, have high angular completeness, and be volume
limited. Three-dimensional void construction may be severely impaired by the presence of a
significantly varying angular selection function on the sky, and similarly by systematic
variation in the number density with redshift. We also require a point distribution that occupies a
large volume, to generate a statistical sample that can be useful for cosmology and also to capture
the morphology of the largest voids. Given these requirements, the SDSS I+II DR7 Main Galaxy catalog is
ideally suited for our purpose. Specifically, we adopt the Korea Institute for Advanced Study Value
Added Galaxy Catalog \citep[KIAS-VAGC;][]{Choi:2007}, which is based on the original catalog
\citep{Blanton:2005} but is supplemented with additional redshift information from the Zwicky
catalog \citep{Falco:1999}, the IRAS Point Source Catalog Redshift Survey \citep{Saunders:2000}, the
Third Reference Catalog of Bright Galaxies \citep{deV:1991} and the Two-Degree Field Galaxy Redshift
Survey \citep{Colless:2001}, together with the updated angular selection function. 

\begin{figure}
\includegraphics[width=0.98\textwidth]{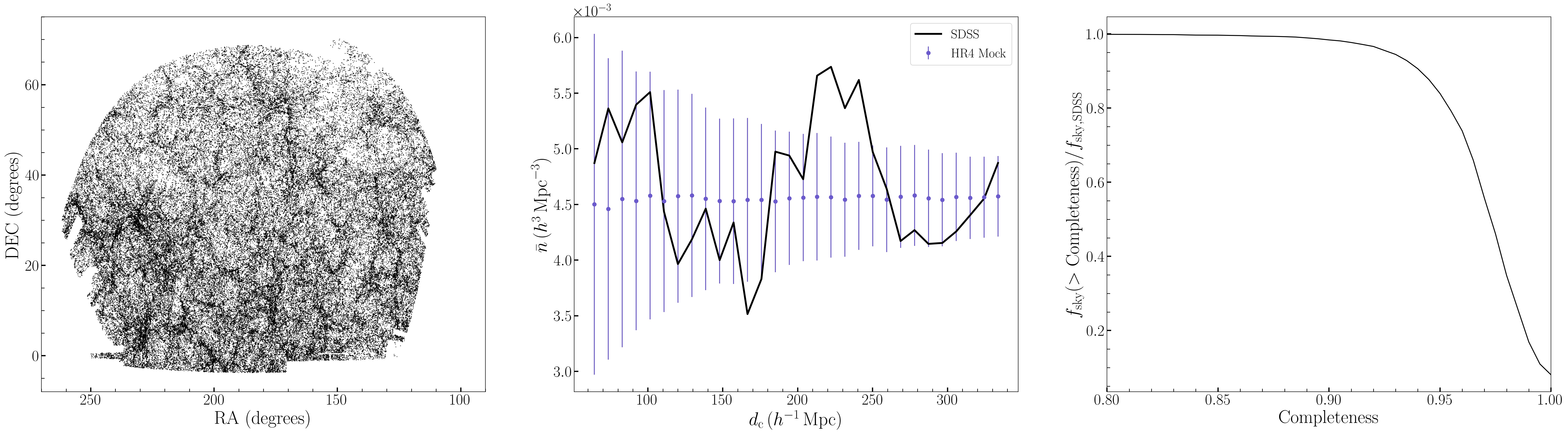}
\caption{\label{fig:1} The SDSS galaxy point distribution as a function of angular coordinates on the sky (left panel). The number density $\bar{n}$ of the SDSS volume limited sample as a function of comoving distance is presented in the middle panel (solid black line), along with the mean and RMS values of the mocks (blue points/errorbars). In the right panel, we present the fraction of the SDSS footprint as a function of its angular completeness. approximately $90\%$ of the footprint has completeness greater than $0.95$. }
\end{figure}

The catalog contains $593,514$ redshifts, with $r$-band Petrosian magnitudes in the range $10 <
r_{\rm p} < 17.6$. Since we require contiguous data, we remove the three southern stripes and the
Hubble Deep Field region. Approximately $90\%$ of the angular mask has completeness higher than $95\%$.
In Figure \ref{fig:1} (right panel) we present the completeness of the angular selection function as
a fraction of the area occupied on the sky. We also present the full galaxy catalog in angular
coordinates (right ascension vs declination, left panel). The redshift and corresponding $r$-band
absolute magnitude cuts are $0.02 \leq z \leq 0.116$ and $M_{r} \leq -20.19$ respectively\footnote{The term $+5\log{h}$  is dropped, i.e., $h=1$ is used, in the conversion from apparent magnitudes to absolute magnitudes.} for the volume
limited subsample that we select for our analysis. The evolution correction for the absolute magnitudes, $E(z)=1.6(z-1)$, is applied \citep{Tegmark:2004}. In the middle panel of Figure \ref{fig:1}, we
present the number density of this volume limited sample as a function of comoving distance,
assuming a flat $\Lambda$CDM cosmology with $\Omega_{m} = 0.3$. The black solid line is the number
density of the SDSS data, and the blue points/errorbars are the mean and rms of the mocks, described
in the following section. 

For the purposes of our study, the SDSS galaxies are not weighted by the angular selection function at any point during the persistent homology calculation and subsequent void finding algorithm. Because the completeness of the value added catalog is so high in this particular data set, we do not expect the angular selection function to have any significant effect on the resulting properties of the voids. This may be an issue for higher redshift data, where the completeness on the sky is generally lower and data quality is worse.  


\subsection{Mock Galaxy Data} \label{sec:data_mock}

We compare the void catalog from the SDSS data to mock catalogs designed to match the clustering
properties of the observed galaxy distribution. For this purpose, we adopt the Horizon Run 4 (HR4)
simulation. HR4 is a cosmological scale $N$-body simulation in which $6300^{3}$ cold dark matter particles
were gravitationally evolved in a $V = (3150 \, h^{-1} \, {\rm Mpc})^{3}$ box using a modified
version of GOTPM code\footnote{The original GOTPM code is introduced in \citet{Dubinski:2004}. A
review of the modifications introduced in the Horizon Run project can be found at
\href{https://astro.kias.re.kr/~kjhan/GOTPM/index.html}{https://astro.kias.re.kr/\textasciitilde
kjhan/GOTPM/index.html}}. 
The WMAP5 cosmology \citep{WMAP5} is used for this simulation.
Details of the simulation can be found in \citet{Kim:2015}. Dark matter
halos and subsequently mock galaxy catalogs are constructed in \citet{Hong:2016} using a most-bound
halo particle-galaxy correspondence algorithm, with satellite galaxy survival time after mergers
calculated using a modified model of \citet{Jiang:2007}

\begin{equation} {t_{\rm merge} \over t_{\rm dyn}} = {\left(0.94\epsilon^{0.6} + 0.6 \right)/0.86
\over \ln \left[ 1 + (M_{\rm host}/M_{\rm sat})\right]} \left({M_{\rm host} \over M_{\rm sat}}
\right)^{1.5} , \end{equation} 

  \noindent where $t_{\rm dyn}$ is a dynamical timescale -- the orbital period of a virialized
  object -- $M_{\rm host}, M_{\rm sat}$ are the host halo and satellite masses, and $\epsilon$ is
  the circularity of the satellite's orbit at the time of merger. The particle and halo mass resolutions of the simulation
  are $9.0\times10^9 \, h^{-1} \, M_\odot$ and $2.7\times10^{11} \, h^{-1} \, M_\odot$, respectively. 

We use the $z=0$ snapshot box to generate mock catalogs -- $N_{\rm r} = 360$ observers are placed in
the box, maximally separated to ensure the mock data does not overlap. 
To reproduce the survey geometry of the SDSS data we use, the SDSS angular footprint is applied relative to each observer placed at the corners. 
A global mass cut is
applied to all galaxies in the snapshot box to ensure that the average number density $\bar{n}$ matches that of the SDSS data,
$\bar{n} = 4.6 \times 10^{-3} \, (h^{3} \, {\rm Mpc}^{-3})$. For absolute magnitude-limited catalogs with a high magnitude selection function like the one used in this work, simulated data with a mass cut typically provides a very good match to a magnitude cut.

The galaxy positions that we obtain from spectroscopic redshifts are shifted by Doppler effect due to the radial component of their own motions. This effect is called the redshift space distortions \citep{Kaiser:1987}, and the modulation from the real-space position ${\bf r}$ to the redshift-space (observed) position ${\bf s}$ caused by the galaxy velocity ${\bf v}$ is expressed as
\begin{equation}\label{eq:rsp} {\bf s} = {\bf r} + {1 \over aH} {\bf \hat{e}}_{\parallel}  ({\bf v} \cdot {\bf \hat{e}}_{\parallel}),
\end{equation} 
where $a$, $H$, and ${\bf \hat{e}}_{\parallel}$ are the scale factor, Hubble parameter, and the unit vector along the joining line between a galaxy and the observer. We apply this correction to each simulated galaxy to generate the redshift-space mock catalogs that are consistent with the observation data. We only perform the PH analysis for the redshift-space mocks and do not repeat it for the real space as the purpose of mocks is to check the consistency between observational data and the predictions from the $\Lambda$CDM cosmology simulation. Studying the difference in void statistics for these two spaces is beyond the scope of the present study and will be pursued in future works.

%

\section{Results} \label{sec:results}

\subsection{Topologically significant voids computed for SDSS DR7 and mocks exhibit similar properties}

\begin{figure}[tbhp!] \centering \includegraphics[width=0.85\textwidth]{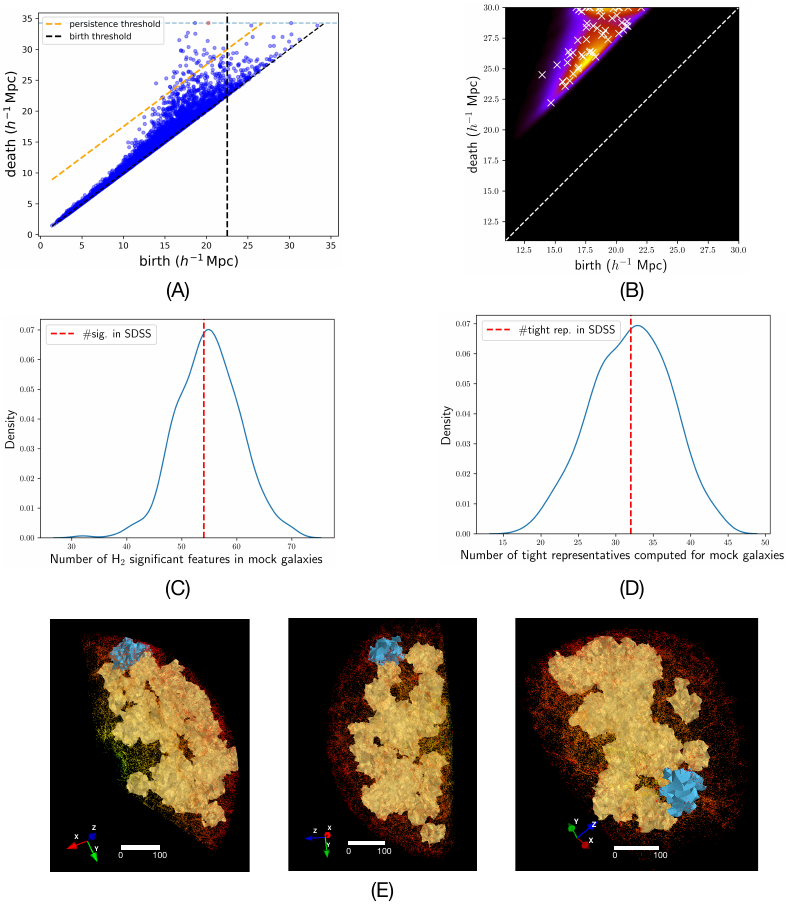}

  \caption{\label{fig:results1}Topology of the mocks agrees with SDSS. (A) H$_2$ PD for SDSS DR7
  computed till a threshold of $30 \, h^{-1} \, {\rm Mpc}$. Significant features are defined as
  those above the orange dashed line (persistence threshold of $7.5 \, h^{-1} \, {\rm Mpc}$) and to
  the left of the black dashed line (birth threshold of $22.5 \, h^{-1} \, {\rm Mpc}$). (B) KDE plot
  of significant features over all mocks. Significant features in SDSS are shown by white x markers.
  (C) Number of significant H$_2$ features. (D) Distribution of numbers of computed tight H$_2$
  representatives in mocks. (E) Three views of computed tight H$_2$ representatives in SDSS. One of
  the voids is highlighted in blue color.}

\end{figure}

We compute PH of SDSS DR7 for Vietoris-Rips filtration (VR-filtration, see Appendix
\ref{app:pers_back}) up to a threshold for the spatial scale of $36 \, h^{-1} \, {\rm Mpc}.$
Figure~\ref{fig:results1}A shows the resulting H$_2$ PD, and in particular presents an important characteristic of the cosmic web -- it is a multi-scale phenomenon. Voids are born, die and exhibit persistence thresholds over the entire range of scales that can be reasonably probed by the SDSS data, from $1 \, h^{-1} \, {\rm Mpc}$ to $50 \, h^{-1} \, {\rm Mpc}$. 
We define a feature to be significant if it has persistence at least $7.5 \, h^{-1}$ Mpc (persistence threshold) and is born before $22.5 \, h^{-1}$ Mpc (birth threshold). We explain these choices in detail in Appendix \ref{sec:app_choose_thresholds}.
Subsequently, there are $57$ significant topological features in SDSS. Based on
our choice of thresholds, it suffices to compute PH up to $30 \, h^{-1} \, {\rm Mpc}$ to determine
the significant features in all of the mocks. This reduces computational run time by more than a
factor of three as compared to the computation up to $36 \, h^{-1} \, {\rm Mpc}$. Computation costs for computing PH up to $30 \, h^{-1} \, {\rm Mpc}$ and representative boundaries of non-trivial features are shown in Appendix \ref{sec:supp_computation_cost}. With our choice, we are selecting objects that are present at scales $\gtrsim {\cal O}(10 \, h^{-1} {\rm Mpc})$, which are those typically parsed by the cosmology community.

Figure~\ref{fig:results1}B shows a kernel density estimation (KDE) plot of distributions of births and
deaths of significant features in all mocks, along with those in SDSS (white x markers).  The
overlay shows that H$_2$ topology is similar across mocks and SDSS. Figure~\ref{fig:results1}C shows
that the number of significant features in SDSS agrees with the median of the number of significant
features in mocks. A total of $32$ tight representatives around single significant voids were
computed for SDSS. The number of tight representative boundaries is less than that of significant
features due to generally larger birth scales and unchanged death scales after the shortening
procedure.
We explain this in Appendix \ref{app:shortening}. Figure~\ref{fig:results1}D shows the distribution of
the number of tight representatives around single voids computed for mocks.
Figure~\ref{fig:results1}E shows three different views of all the representative boundaries computed
in SDSS. We visually observe that the computed boundaries are polyhedral and not necessarily convex.

\begin{figure}[tbhp!]
\centering
\includegraphics[scale=0.7]{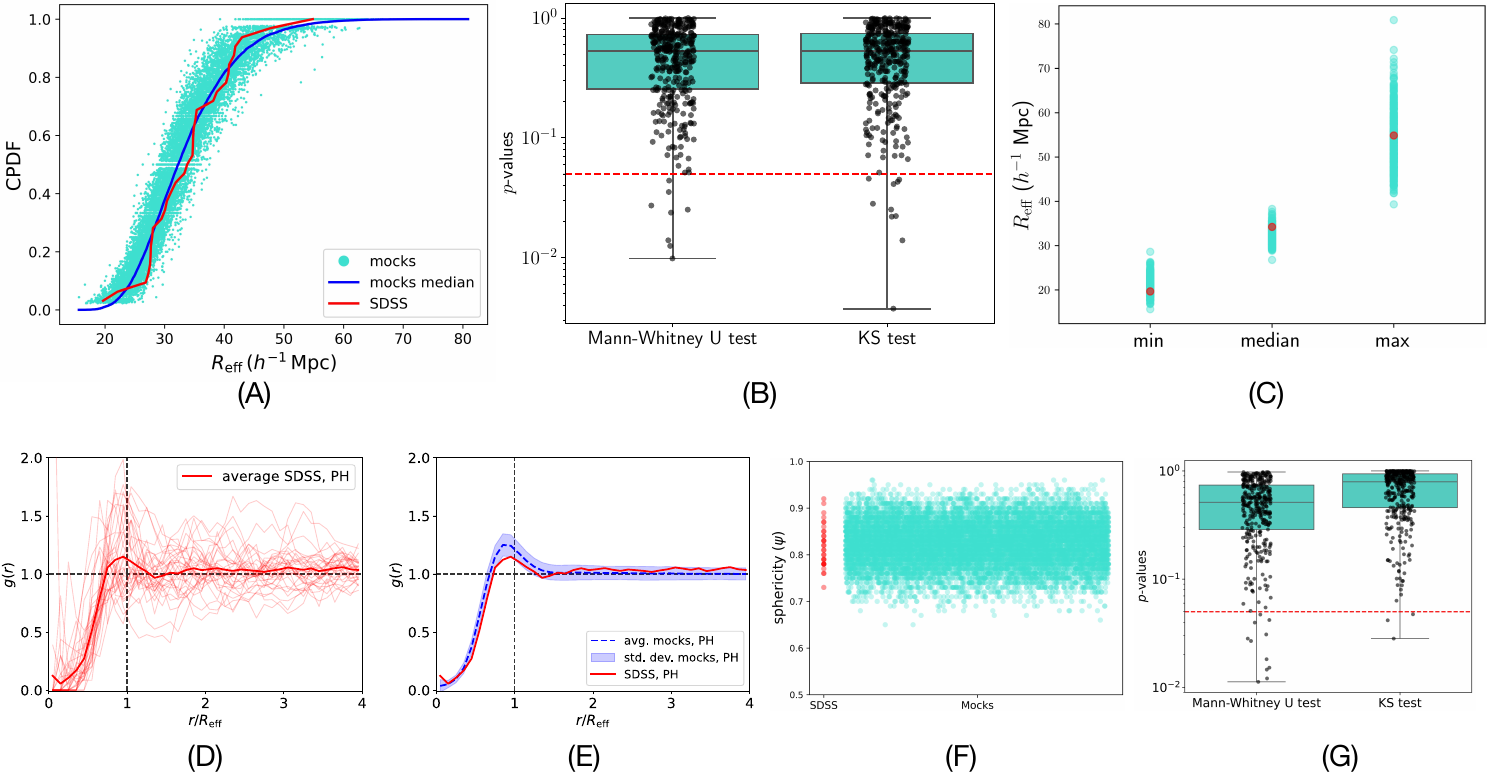}

\caption{\label{fig:results2} SDSS and the 360 mocks exhibit similar morphology. (A) Cumulative PDF
  (CPDF) of $R_{\rm eff}$ for SDSS (red) and all mocks (turquoise). (B) $p$-values from Mann-Whitney
  U and KS significance tests comparing the distribution of $R_{\rm eff}$ of voids in SDSS to the
  distributions of $R_{\rm eff}$ of voids in the mocks. Majority are greater than $0.05$ (red dashed
  line). (C) Minimums, medians, and maximums of distributions of effective radii of voids computed
  for mocks (turquoise) and SDSS (red). (D) Radial distribution functions of each void that is
  computed for SDSS. The average of radial distributions over all voids is shown in a bold red line
  plot. (E) Averages of radial distribution functions of voids computed for mocks follow similar
  pattern to the average of the voids in SDSS. (F) $p$-values from Mann-Whitney U and KS
  significance tests show that the sphericities of voids is less than $1$ but greater than $0.6.$
  (G) $p$-values from comparing distributions of sphericities of voids in mocks to that in SDSS. }

\end{figure}

We compute the effective radius ($R_{\text{eff}}$) of each void as the radius of the sphere with the
same volume as the convex hull of its computed representative boundary. Cumulative probability
density functions (CPDF) of $R_{\rm eff}$ distribution of all voids in SDSS is compared to that of
mocks in Figure~\ref{fig:results2}A. The distribution of $R_{\rm eff}$ of the voids computed for
SDSS is compared to that of each of the $360$ mocks using Mann-Whitney U test and Kolmogorov-Smirnov
(KS) test.  Figure~\ref{fig:results2}B shows that the distribution of $R_{\rm eff}$ of the voids in
the majority of the mocks is not significantly different from that of SDSS ($p$-value $\gg 0.05$).

Figure~\ref{fig:results2}C shows that the minimums of the effective radii computed for the mocks
range from approximately $15$ to $30 \, h^{-1} \, {\rm Mpc}$ and the medians range from $30$ to $40
\, h^{-1} \, {\rm Mpc}$, matching closely to SDSS. The maximums for the mocks, however, show a wider
range of, approximately, $40$ to $80 \, h^{-1} \, {\rm Mpc}$.

Since computed H$_2$ representatives are polyhedral that might be very different from spherical
approximations of boundaries of voids, for additional validation we compute the radial
distribution function $g(r)$ for each void with the centroid of its computed representative boundary
as the reference point. Briefly, $g(r)$ is the ratio of density of shells of thickness $2 \, h^{-1}
\, {\rm Mpc}$ around the centroid of the void to the density of shells containing a random point distribution 
($\approx$ $10^6$ points) distributed uniformly in the SDSS footprint. It is normalized by the mean value of the mocks at $r=4R_{\rm eff}$, and the radii of the shells are finally normalized by $R_{\rm eff}$. Hence, $g(r)>1$($g(r)<1$) indicates overdense(underdense) regions. This ratio accounts for the fact that at large distances from the centroid location, the shells will hit the boundary of the survey. So, the density of the galaxies will artificially drop but so will the density of random points. 

Figure~\ref{fig:results2}D shows the $g(r)$ profiles of each void in SDSS for which a
representative boundary was computed using PH. On average, $g(r)$ rises for $r< R_{\rm eff}$
with peak very close to $r = R_{\rm eff}$. Figure~\ref{fig:results2}E shows that the averages of
$g(r)$ profiles of all mocks present a similar profile to the average of the SDSS. We note that the peak
is not exactly at $r=R_{\rm eff}$ presumably because the voids in the universe are not
exactly spherical. We quantify this by computing sphericity ($\psi$) as the ratio of $4\pi R_{\rm
eff}^2$ to the surface area of the computed polyhedral boundary. Figure~\ref{fig:results2}F shows
that the sphericity of all voids is less than $1$ but more than $0.6.$ The average sphericity of the
voids is around $0.8$ in every configuration (SDSS and the 360 mocks). The distribution of the
sphericities of the voids in the 360 mocks is compared to the distribution of those in SDSS using
the Mann-Whitney U test and the Kolmogorov-Smirnov (KS) test. Figure~\ref{fig:results2}G shows that the
majority of mocks do not show significant differences ($p$-value $>0.05$) in sphericities of the
voids as compared to SDSS. We expect voids to be aspherical because of the complex morphological structure of the cosmic web, and also due to intrinsic anisotropic observational artifacts such as redshift space distortion and the Alcock-Paczynski effect.

\begin{figure}[tbhp!]
\centering
\includegraphics[width=\textwidth]{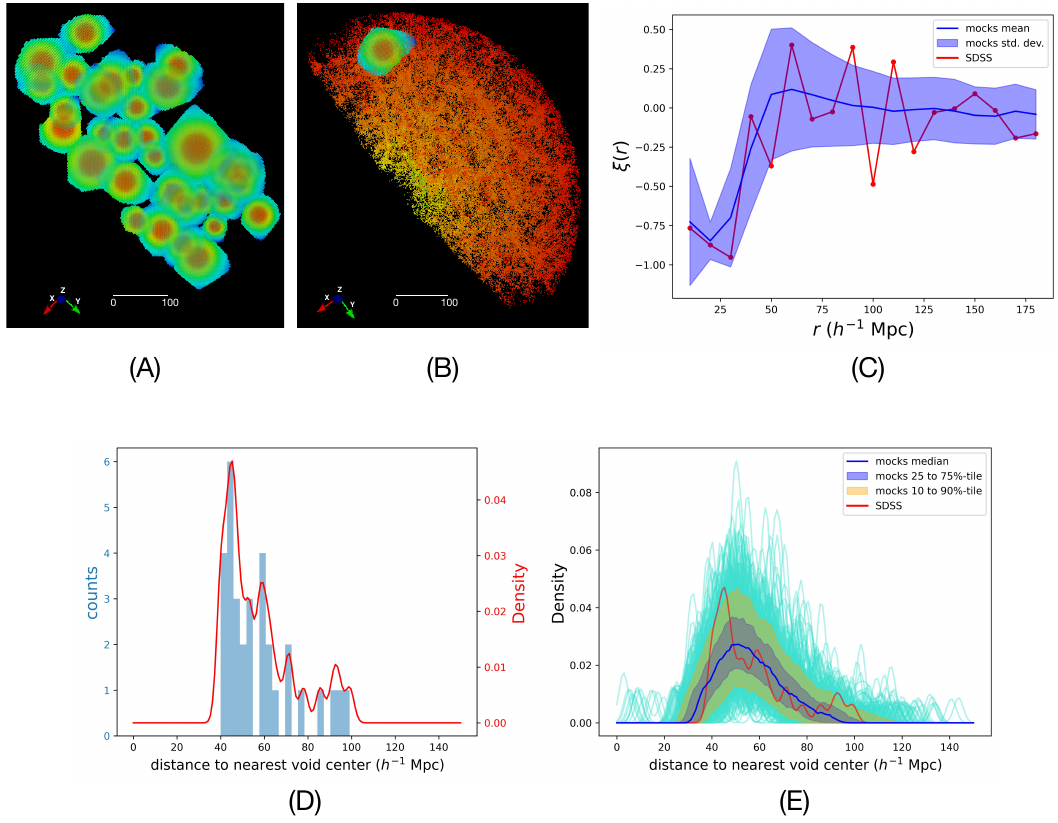}

\caption{\label{fig:results3}Characteristic distances between nearest void centers. (A) All cubic voxels that are inside
  voids computed for SDSS. (B) An example
  of cubic voxels inside one of the computed voids in SDSS. The color scheme is scaled based on the distance of voxels from the
  centroid of the void it is in. (C) The two-point correlation function of the void central positions. The red line represents the SDSS data, while the blue line and area indicate the mean and standard deviation from mocks. (D) Distribution of distances of the nearest void
  center to all void centers. The KDE of the distribution is shown in red. (E) KDE of the
  distributions of distances of the nearest void center to all void centers for the mocks
  (turquoise). The median of the KDEs for mocks is shown in blue.}

\end{figure}

Figures \ref{fig:results3}A and \ref{fig:results3}B show the PH voids computed in SDSS. For each void center, we
compute the distance of the nearest void center from it. 
Figure \ref{fig:results3}C shows the two-point correlation function of void centers. We use the Landy-Szalay estimator \citep{1993ApJ...412...64L}, $\xi(r)=\frac{DD-2DR+RR}{RR}$, where $DD$ takes the pair count of void centers in separation bins $[r-\Delta r,r+\Delta r]$, $RR$ is the pair count of random points distributed within the SDSS survey volume, and $DR$ is the cross pair count between void centers and randoms. The number of random points is set to be sufficiently large, and $DR$ and $RR$ are normalized to $DD$ accordingly. We observe that $\xi(r)$ is peaked at $\sim 50$--$60\; h^{-1}{\rm Mpc}$ and approaches zero as separation $r$ increases. This means that the separation of voids is typically $\sim 60 h^{-1}; {\rm Mpc}$, which is also observed in Figure \ref{fig:results3}D. The drop below $50\; h^{-1}{\rm Mpc}$ is due to the size of voids, i.e., the exclusion of other voids \citep{Hamaus:2013qja, 10.1093/mnras/stv2973, Shim:2020qav}.
Figure \ref{fig:results3}D shows the
distribution of nearest neighbor distances for the SDSS. The KDE of this distribution is shown in red; the majority of nearest void centers are within $40$ to $60 \, h^{-1} \, {\rm Mpc}.$ Figure
\ref{fig:results3}E shows KDEs of similar distributions for all of the mocks (turquoise).  Although noisy due to a low number of objects in each realization, we find that the peak in the median of the KDEs of mocks (blue) matches with the peaks in the KDE of the
SDSS (red).

The measured two-point statistics of the void centers -- nearest neighbor separation and correlation function -- present behaviour that is typical of critical points \citep{10.1093/mnras/238.2.293}. The anti-clustering (`exclusion') regime of the correlation function  $r < 50 \, h^{-1} \, {\rm Mpc}$ is related to our choice of birth and persistence thresholds because these indirectly determine the comoving size of the voids. In terms of critical points of some underlying continuous field traced by the galaxies, the exclusion zone can be understood as the typical length scale required for the curvature of the field to change signs twice between adjacent voids \citep{Shim:2020qav}.

\subsection{Comparison with Other SDSS Void Catalogs}\label{sec:comparison}

\cite{douglass2023updated} compute and compare void catalogs for SDSS DR7 using two popular classes
of void finding algorithms, VoidFinder~\citep{el1997voids} and V$^2$~\citep{neyrinck2008zobov}. The
former computes voids by first expanding spheres centered at locally empty regions till they are
bounded by a threshold number of galaxies, and these spheres are then merged to define
voids~\citep{douglass2023updated}. The latter first computes a 3D Voronoi tessellation of the
distribution of galaxies, then combines computed Voronoi cells into zones using watershed
segmentation, which are finally merged into voids~\citep{douglass2023updated}. Since VoidFinder does
not compute voids with an effective radius larger than $30 \, h^{-1} \, {\rm Mpc}$ \citep{douglass2023updated} and
PH computes many larger voids, we only compute V$^2$ voids in the adopted KIAS-VAGC in this study for
comparison with the computed PH voids.

\begin{figure}[tbhp!] \centering
  \includegraphics[scale=0.75]{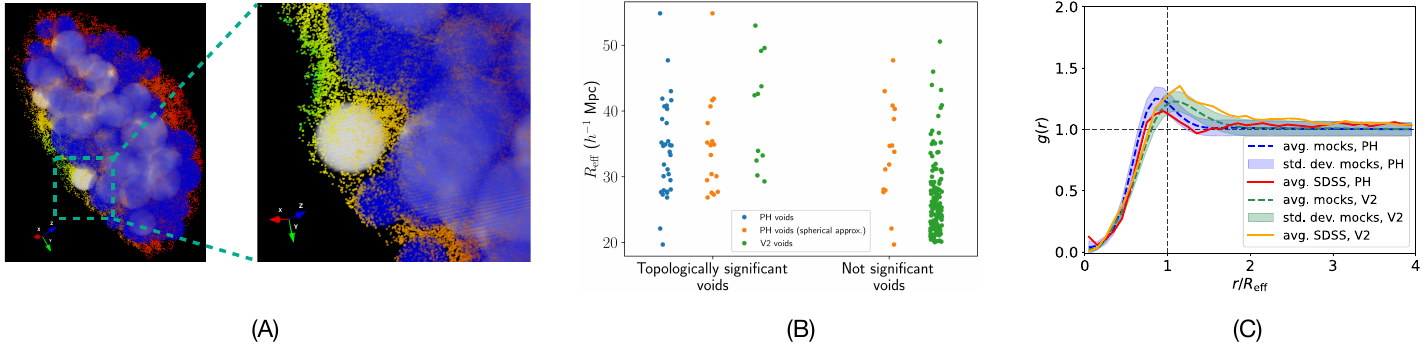}

  \caption{\label{fig:results4}Comparing PH and V$^2$ voids. (A) PH voids in white and V$^2$ voids
  in blue. PH finds a void not found by V$^2.$ (B) Distributions of $R_{\rm eff}$ of V$^2$ and PH voids, categorized on the basis of whether their representative boundaries contain topologically significant features. Majority of V$^2$ voids do not contain topologically significant features. (C) $g(r)$ of PH and V$^2$ for SDSS and mocks.}

\end{figure}

We compute V$^2$ voids using VAST toolbox \citep{douglass2022vast} with settings $H_0=100 \, h \,
{\rm km} \, {\rm s}^{-1} \, {\rm Mpc}^{-1}$ and $\Omega_m=0.3.$ A total of $419$ voids are reported
with effective radius at least $10 \, h^{-1} \, {\rm Mpc}.$ We consider voids with $R_{\rm eff}$ at
least $20 \, h^{-1} \, {\rm Mpc},$ resulting in $150$ voids. Figure~\ref{fig:results4}A shows spherical approximations of PH
voids and V$^2$ voids based on $R_{\rm eff}.$ The zoomed-in panel on the right in the figure shows a PH
void that is not found by V$^2.$ 

We first compare the number of topologically significant voids. We say that a PH void is topologically significant if the smallest hyperrectangle around its representative boundary contains at least one significant feature. For a $V^2$ void, we compute the number of  significant features in the smallest hyperrectangle around the sphere of the void's effective radius centered at its centroid. For a fair comparison, we do the same for every PH void, and these are labeled as `spherical approximation'. Figure~\ref{fig:results4}B shows that only $11$ out of the $150$  $V^2$  voids (green points) are topologically significant. On the other hand, all of the PH voids are topologically significant (blue points). Only around half of the spherical approximations of PH voids do not contain a significant feature. However, they are still more than the number of topologically significant $V^2$ voids. We further note that the discrepancy between the number of significant PH voids and significant spherical approximations of PH voids indicates that some of the PH voids are aspherical.
Finally, we compare $g(r)$ profiles for PH and V$^2$ voids in the SDSS and mocks in Figure~\ref{fig:results4}C. We note that the peak of $g(r)$ for PH voids occurs at $r < R_{\rm eff}$
(at $\frac{r}{R_{\rm eff}}=0.95)$ whereas for V$^2$ voids it occurs at $r > R_{\rm eff}$
(at $\frac{r}{R_{\rm eff}}=1.15).$ This indicates that voids inferred using the two different methodologies possess different morphologies.

\section{Discussion} \label{sec:disc}

We have introduced a void finding algorithm that is based on the rigorous mathematics of persistent
homology, and applied it to the SDSS DR7 main galaxy sample and a set of mock catalogs constructed
from a cosmological scale dark matter simulation. Certain user inputs are required to define voids,
and in this work a birth threshold of $\tau_{u} = 22.5 \, h^{-1} \, {\rm Mpc}$ and persistence level
$p = 7.5 \, h^{-1} \, {\rm Mpc}$ are used to select topological robust objects from the galaxy point
distribution. We find representative boundaries around $32$ unique voids that satisfy the criteria
imposed, comprising a total volume fraction of $0.26$ of the SDSS footprint over the redshift range
$0.02 \leq z \leq 0.116$.  The physical properties of the voids have been ascertained; chiefly their
radial profiles, effective radii distributions, the nearest neighbor separation, two-point correlation function and sphericity. We
find a range of sizes between $21-56 \, h^{-1} \, {\rm Mpc}$, and a median nearest neighbor
separation of $\sim 57 \, h^{-1} \, {\rm Mpc}$.  The properties of these objects will depend on the
choice of birth threshold and persistence. The SDSS voids show excellent agreement with the same
quantities extracted from the mock catalogs, indicating that the large scale distribution of matter
in the observed Universe closely match our expectations from simple cold dark matter gravitational
physics. This is a non-trivial result -- the mock galaxies have been selected to match the two point statistics of the SDSS, but the spatial distribution and morphological properties of voids carry information beyond the power spectrum. The void profiles; $g(r)$, obtained using persistent homology and the V$^{2}$ algorithm are in reasonable agreement, indicating that the void profile is a robust statistical quantity that can be used for cosmological parameter estimation. However, the peak of $g(r)$ occurs at mildly different $r$ values for the two algorithms, indicating that the non-spherical nature of the voids will play some role in determining the average shape.

The topological objects defined in this work are constructed from the observed point distribution rather than a smoothed density field inferred from the galaxy positions. This makes it difficult to analytically relate the void properties to cosmological parameters, since the standard cosmological model is predicated on a fluid description of matter. However, there are a number of interesting avenues that remain to be explored. First, the sensitivity of the void properties to galaxy bias, and cosmological parameters, can be tested by applying our algorithm to other mock galaxy and dark matter data. We expect only mild dependence on galaxy bias, since all galaxies should trace the same wall-like structures on large scales. The cosmological parameters $\Omega_{\rm m}h^{2}$ and $n_{\rm s}$ will determine the extent to which the dark matter field fluctuates, and the void statistics may be sensitive to these quantities. Relating the topology of the point distribution to that of the smoothed density field is also an on-going point of interest. Persistence provides a way of divining the significance of features found using persistent homology. Alternatively, the smoothing scale used to convert a point distribution into a continuous field washes out small scale holes and provides a measure of significance based on physical scale. In addition, smoothing with a Gaussian kernel will sphericalize voids that are roughly equal in size to the smoothing volume, an effect that will not be present in the voids inferred from the point distribution. The fluid and particle descriptions will only match on large scales, for objects significantly above the smoothing scale used to define the fluid. Relating the properties of voids obtained using the fluid and particle descriptions of matter would provide a link between the two different interpretations of the density field that generates spacetime curvature. 

The topology of cosmological fields; the late time galaxy distribution, Cosmic Microwave Background, weak lensing maps, contains information beyond summary statistics that are commonly used by the community. Extracting this information, and comparing the results to mock data, provides an important consistency check of the $\Lambda$CDM model. Going further and inferring the ensemble average of topological summary statistics of random fields remains a long-standing goal.

\section*{Acknowledgements}
We warmly thank Juhan Kim and Sungwook E. Hong for generating the Horizon Run 4 mock galaxy catalogue used in this work. MT and SA are supported by an appointment to the JRG Program at the APCTP through the Science and Technology Promotion Fund and Lottery Fund of the Korean Government, and were also supported by the Korean Local Governments in Gyeongsangbuk-do Province and Pohang City. SA also acknowledges support from the NRF of Korea (Grant No. NRF-2022R1F1A1061590) funded by the Korean Government (MSIT). MT was supported by the National Research Foundation of Korea (NRF) grant funded by the Korea government (MSIT) (2022R1F1A1064313). We thank the Korea Institute for Advanced Study for providing computing resources (KIAS Center for Advanced Computation Linux Cluster System). A large data transfer was supported by KREONET, which is managed and operated by KISTI.

\appendix

\section{Persistent homology background and terminology}\label{app:pers_back}

\begin{figure}
\centering
\includegraphics[width=0.85\textwidth]{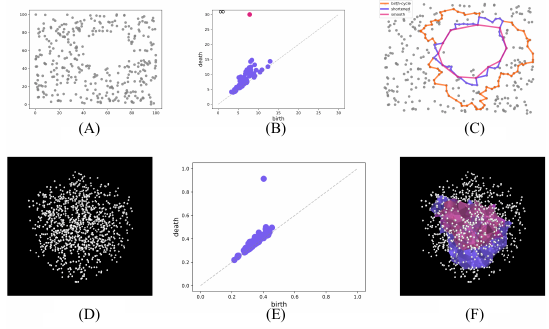} 

\caption{\label{fig:fig1}(A) A noisy discrete data set in with one significant hole. (B) The births
  and death of holes are plotted as a persistence diagram. This example shows that only one hole has
  relatively high persistence. (C) Representative boundaries tighten as they are shortened by our
  algorithm.  (D) A point-cloud in 3D. No significant hole is visible. (E) PD shows that one feature
  has relatively high persistence. (F) Our algorithms improve geometric precision. The blue boundary
  is the birth-cycle from the recursive algorithm. The red boundary is the smooth cycle after
  applying the greedy shortening and smoothing algorithms.}

\end{figure}

Figure \ref{fig:fig1}A shows an example of a discrete set of points or a point-cloud.  It has many
gaps and holes, but one stands out as a larger hole compared to others. Dimension-$1$ persistent
homology (PH) computes the birth and death of holes across different spatial scales. This
information is plotted as a persistence diagram (PD). Figure \ref{fig:fig1}B shows dimension-$1$ PD
computed for the example in Figure \ref{fig:fig1}A. The persistence of a hole is defined as the
difference between its death and birth. Those with higher persistence are more robust to noise in
the data set. There is one feature (marked in red) with relatively higher dimension-$1$ persistence,
as expected. Figure \ref{fig:fig1}C shows multiple possible representative boundaries computed
around the feature with maximum persistence.

%

\section{PH computation}

Locations of topological voids were determined by computing tight representative boundaries of
significant H$_2$ features in VR-filtration of the raw data. Features born at a spatial scale at
most $\tau_u$ and with persistence at least $p$ were defined as significant. These parameters
are user-defined. PH was computed up to threshold of $\tau_u+p$ using
Dory~\citep{aggarwal2024dory}, which implements the matrix reduction algorithm to compute
PH~\citep{carlsson2005computing}. Matrices in Dory are represented using compressed sparse row
format and the number of non-zero entries that can be represented is the upper limit of unsigned
32-bit $\texttt{int}, l = 4294967295.$ However, computing PH for the galaxy data sets in this work
resulted in matrices with more than $l$ non-zero entries. Additionally, the number of non-zero
columns also exceeds $l$ in these computations. Both these issues were resolved by using a new
sparse matrix format that increased the limits on numbers of non-zero elements in a matrix and
non-zero columns to $l^2.$

\section{Computing tight representative boundaries}\label{app:shortening}

Representative boundaries for all topological features were computed and shortened using algorithms
introduced in~\cite{aggarwal2023tight}. Briefly, columns of the matrix of reduction operations are
used as the initial set of representatives. Representative with birth parameters less than the
user-defined birth threshold are greedily shortened by summing (modulus $2$) boundaries pairwise
that result in maximal reduction in lengths.

\begin{figure}
\centering
\includegraphics[width=0.75\textwidth]{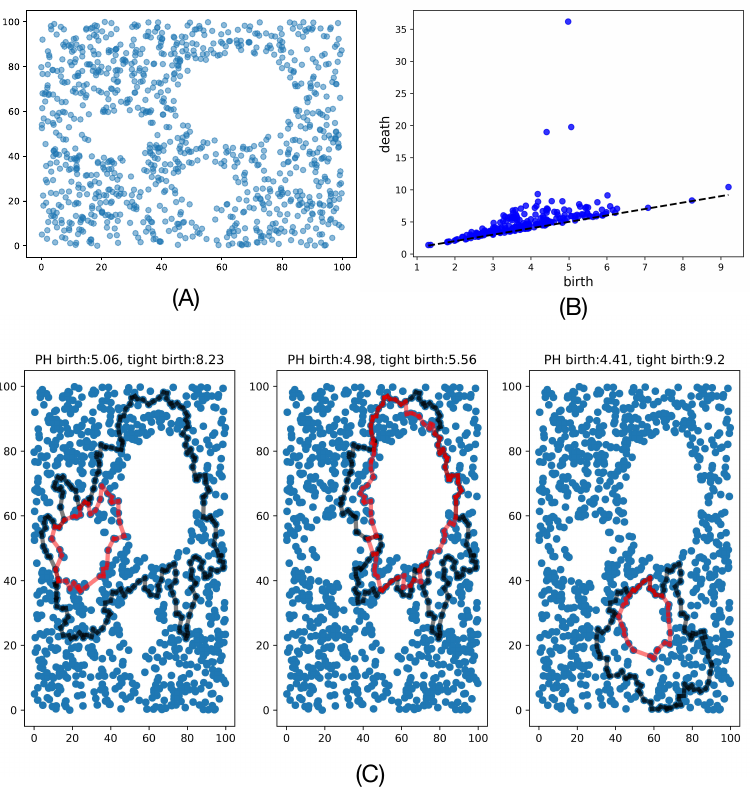} 

\caption{\label{fig:birth_explain} Number of tight representatives can be smaller than the number of
  significant features. (A) A simulated data set embedded in $\mathbb{R}^2$ with three distinct
  holes. (B) H$_1$ PD shows there are three features with relative high persistence. (C)
  Representatives from the matrix reduction algorithm are shown in black. Their birth paramaters are
  $5.06, 4.98$ and, $4.41.$ Tight representatives after greedy shortening ($\tau_u=10$) are shown in
  red. Their birth paramaters are $8.23, 5.56,$ and, $9.2.$ Consequently, they have lower
  persistence and some may be classified as non-significant depending upon the persistence
  threshold.}

\end{figure}

We now explain why the number of computed tight representatives can be less than the number of
significant features. Figure \ref{fig:birth_explain}A shows a simulated data set in two dimensions
with three holes that visually stand out. The H$_1$ PD in Figure \ref{fig:birth_explain}B shows that
there are three features with relatively high persistence. Figure \ref{fig:birth_explain}C shows
representative boundaries computed by the matrix reduction algorithm (black curves) around these
features. The birth parameters of these features are $5.06, 4.98$ and, $4.41.$ Geometrically, the
birth parameter of a feature implies that the algorithm finds a cycle around the feature with the
length of its longest edge equal to the birth parameter. Here we compute tight representatives with
birth threshold $\tau_u=10.$ Representatives with birth parameters less than $10$ are summed modulus
$2$ for greedy shortening. This process results in new representatives that may have the same or
larger birth parameters, but smaller than $\tau_u=10.$ Figure \ref{fig:birth_explain}C shows the
representatives after greedy shortening in red. They are geometrically tighter around the features,
however, their birth parameters are $8.23, 5.56,$ and, $9.2,$ all larger than the birth parameters
of these features from the matrix reduction algorithm ($5.06, 4.98$ and, $4.41,$ respectively).  The
death parameter remains unchanged. As a result, death $-$ birth of tight representatives might be
less than the death $-$ birth of the corresponding features that are computed by the matrix
reduction algorithm. Consequently, the persistence of some of the tight representatives around
significant features may be below the persistence threshold, classifying them as non-significant.
Same reasoning follows for features of higher dimensions.

After greedy shortening, tight representatives that can potentially contain topologically
significant features are stochastically shortened as follows. A local cover of the representative
boundary is computed as the smallest hyperrectangle with points of the boundary inside or on it.
Stochasticity in the cover is introduced by perturbations and permutations as follows. First,
$n_{\text{pert}}$ perturbations (user-defined parameter) are constructed. The maximum perturbation
parameter is initialized as $\Delta_m = p/3.$ For each point $p_i,$ we denote the distance of its nearest neighbor by $d_i.$ Each point $p_i$ is perturbed randomly in a
neighborhood of radius $r_i = \text{min}\{d_i/3, \Delta_m\}.$ $n_{\text{pert}}$ perturbations are
constructed and PH upto $\tau = \tau_u + p$ is computed for each perturbation. If the number of
significant features in each of the $n_{\text{pert}}$ perturbations is not the same as the number of
significant features in the unperturbed cover, we halve $\Delta_m$ and repeat.  Otherwise, we are
done with constructing perturbations of the cover. For each perturbation of a cover, we create
$n_{\text{perm}}$ permutations (user-defined parameter) as follows. A list of edges of lengths up to
$\tau$ is constructed that is sorted by their lengths. The lengths are rounded to decimal precision
$r$ (user-defined parameter). The order of the sorted edges define the order of columns in the
boundary and co-boundary matrices for PH and representative boundary computation.  Permutations are
constructed by permuting the order of edges of the same length. Note that the number of possible
unique permutations might be less than $n_{\text{perm}}.$ PH is computed for (at most)
$n_{\text{pert}} \times n_{\text{perm}}$ sorted lists of edges, for each cover, up to $\tau.$
Representative boundaries born at a spatial scale at most $\tau_u$ are also computed. From all
computations for a cover, a set of all boundaries that possibly wrap around significant feature(s)
is determined. Minimal boundaries are selected from these sets, and they are smoothed using algorithm detailed in S1 Text of \cite{aggarwal2023tight}.

\section{Choosing parameters to define significant features}\label{sec:app_choose_thresholds}

For all data sets in this work, we defined significant features as those born before $\tau_u = 22.5 \, h^{-1}$ Mpc and with persistence at least $p = 7.5 \, h^{-1}$ Mpc. Here, we explain these choices.

\subsection{Persistence threshold}

\begin{figure}
    \centering
    \includegraphics[width=0.5\linewidth]{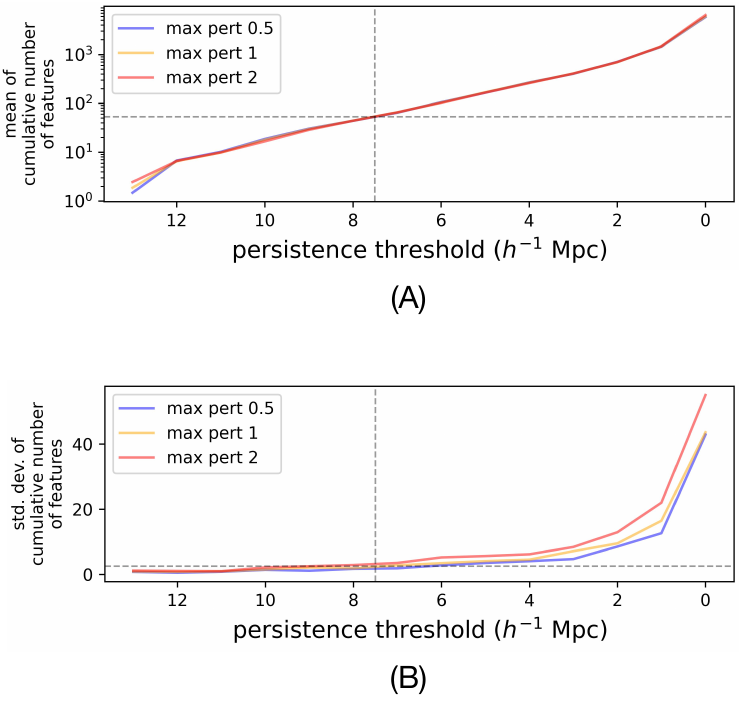}
    \caption{Number of H$_2$ features is computed for $50$ samples of three different perturbations of the data sets. Different colors show different magnitudes of the perturbation. (A) The mean number of features with persistence at most $x$ decreases exponentially as $x$ increases. (B) The standard deviation in the number of features is small and consistent across the three perturbations for persistence thresholds greater than $7.5 \, h^{-1}$ Mpc (dashed vertical line).}
    \label{fig:app_explain_PH_thresh}
\end{figure}

First, we decide the value of persistence threshold. A higher persistence implies higher robustness to noise or perturbations in the data set. Hence, we perturbed location of each galaxy randomly in a ball of radius $\Delta$ and computed PH. This was done for $50$ samples for each of $\Delta=\{0.5, 1, 2\} \, h^{-1}$ Mpc. Then, we count the number of H$_2$ features. Figure \ref{fig:app_explain_PH_thresh}A shows the mean of cumulative counts of  features for the different values of $\Delta.$ For instance, a point $(x_*, y_*)$ on this curve informs that there are $y_*$ non-trivial H$_2$ features (averaged over the $50$ samples) with persistence at most $x_*.$ In other words, the $x$-axis is the persistence threshold. As expected, the number of features increases as persistence threshold decreases (notice that the $x$ axis is reversed in the plot). Figure \ref{fig:app_explain_PH_thresh}B shows the standard deviation in the cumulative number of features. In concordance with the implication of higher persistence, the standard deviation is lower at a higher persistence threshold. Moreover, at low persistence thresholds, standard deviations are inconsistent across the different values of $\Delta.$ At $x=7.5,$ the standard deviation is small and consistent across the different perturbations. At higher thresholds, the number of features decreases exponentially (see Figure \ref{fig:app_explain_PH_thresh}A). Hence, we choose $p = 7.5 \, h^{-1}$ Mpc as a conservative choice for persistence threshold.

\subsection{Birth threshold}

\begin{figure}
    \centering
    \includegraphics[width=1\linewidth]{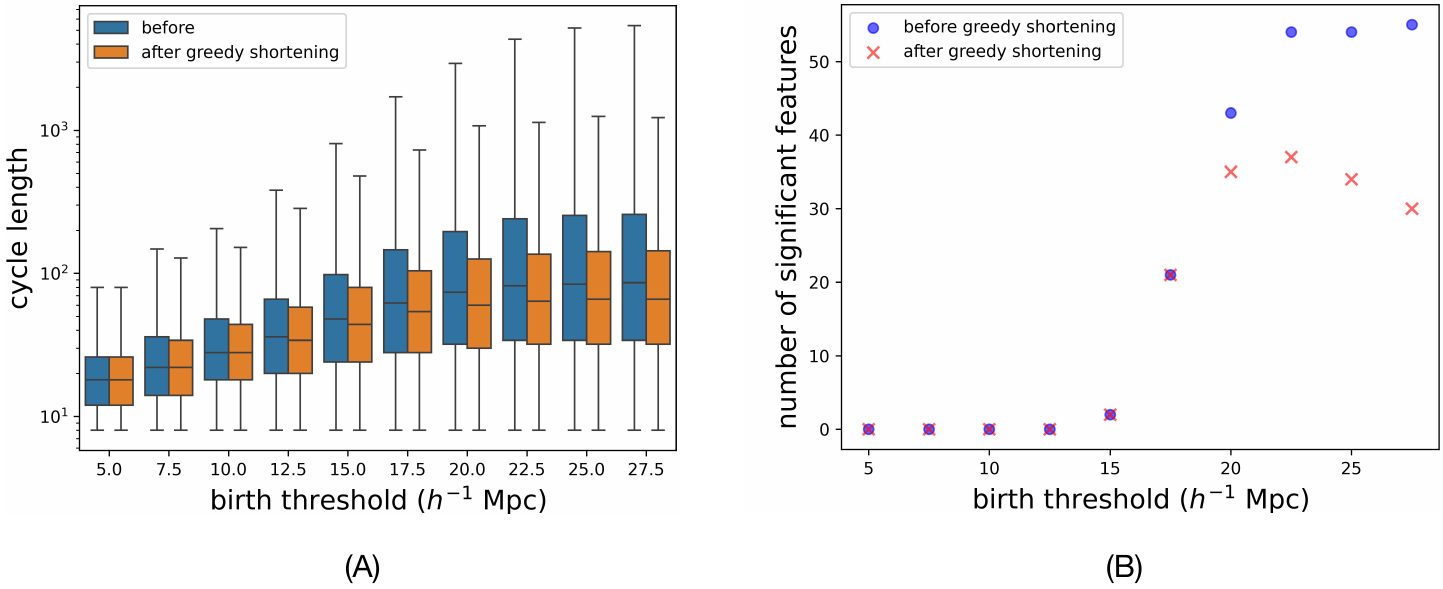}
    \caption{(A) Distributions of cycle lengths before and after greedy shortening at different birth thresholds ($\tau_u$). The distributions are similar for $\tau_u > 17.5 \, h^{-1}$ Mpc. (B) Number of significant features around which we can find a representative after greedy shortening attains maximum at $\tau_u=22.5 \, h^{-1}$ Mpc.}
    \label{fig:app_explain_birth_thresh}
\end{figure}

Given a persistence threshold, we now decide the birth threshold. Figure \ref{fig:app_explain_birth_thresh}A shows  distribution of lengths of representatives, before and after greedy shortening, for different values of birth thresholds. We note that for birth thresholds greater than $22.5 \, h^{-1}$ Mpc, the distributions of cycle lengths after greedy shortening are similar. Further, we compute the number of significant features at each birth threshold. Figure \ref{fig:app_explain_birth_thresh}B shows that the peak in the number of significant features that are enclosed within a shortened representative is at $22.5 \,h^{-1}$ Mpc. Hence, we choose $\tau_u = 22.5 \, h^{-1}$ Mpc as the birth threshold to define significant features.

\section{Sensitivity of Void Properties to Data Variations}

\begin{figure}
\centering
\includegraphics[width=0.4\textwidth]{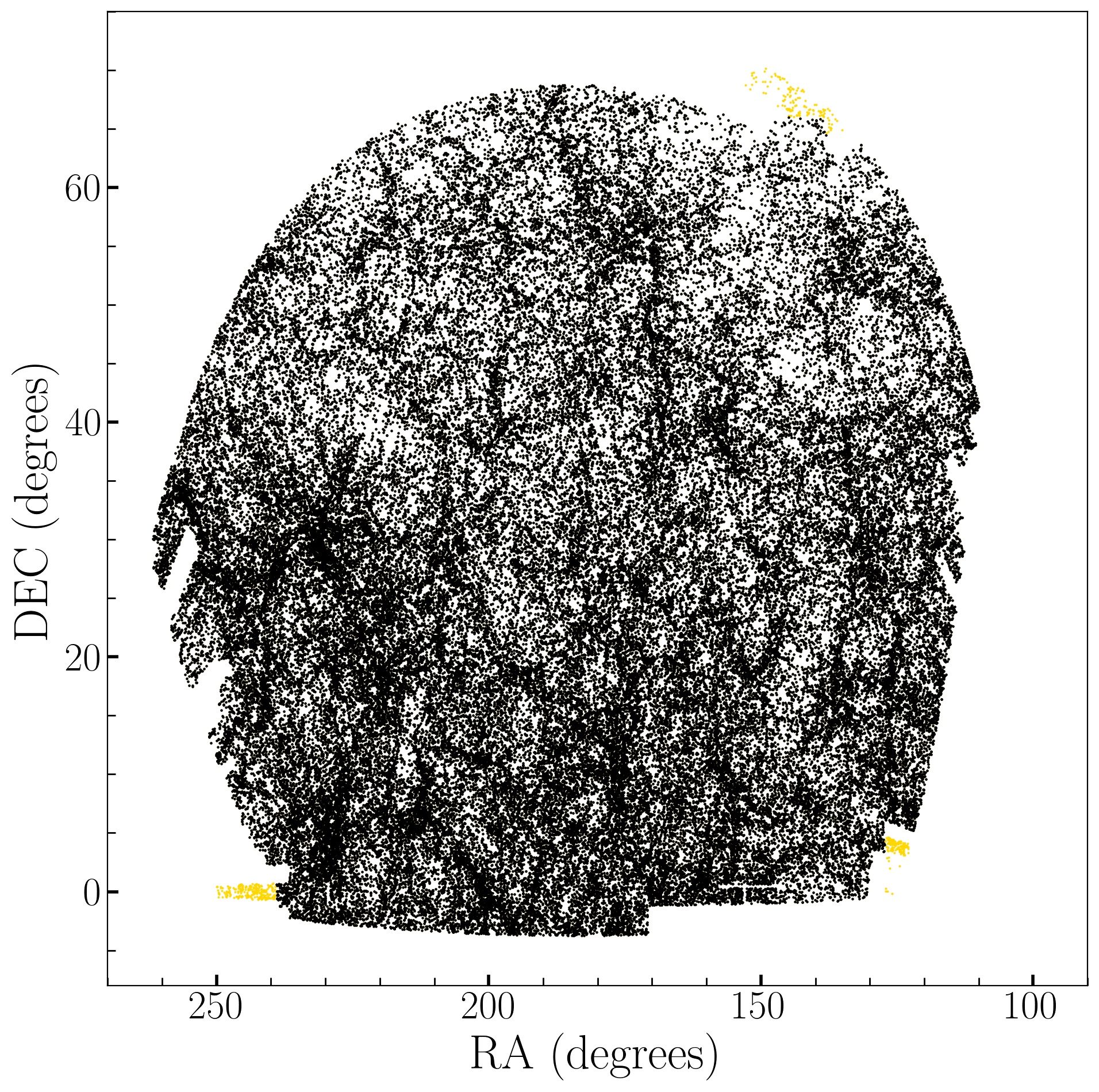} 
\caption{\label{fig:trim} The angular distribution of SDSS galaxies on the sky. In the Appendix, we cut the gold points from the sample and re-analyse the data, to test the sensitivity of the void properties to the boundary. }
\end{figure}

In this section we address two issues. First, we trim a small number of protruding surfaces in the angular footprint of the SDSS. Topological statistics are often particularly sensitive to complicated data boundaries, because the underlying point distribution is inhomogeneously sampled in these regions. Or in the language of smoothed fields, the underlying density field is subject to a larger noise component in the absence of data that is cut by the survey boundaries. We test that the voids found in this work are not significantly compromised by the boundary, by generating a slightly simpler angular footprint. In Figure \ref{fig:trim} we present the full data used in the main body of the paper, but the gold points are cut in this section. In what follows we call the new data set
`trimmed' (cf. black points, Figure \ref{fig:trim}) and the non-trimmed `original'. These particular regions were chosen to maximize the volume to surface area of the data and make the footprint closer to a spherical cap on the sky. If the data is bounded by a complicated mask geometry, then spurious triangulated meshes can be generated that cross the masked regions.

Second, we analyze robustness of the voids found by undersampling both original and trimmed data
sets. We compute and compare PH and tight representatives of original and trimmed data sets that are
undersampled at three different sampling percentages of $p=95, 90$ and,  $80\%$ (three samples each). We call original and trimmed data sets as catalogs. Then, comparison of catalog1-catalog2 at percentages $p_1$-$p_2$ is the Mann-Whitney U rank test of distributions of centroids of a sample of catalog1 at percentage $p_1$ with that of a sample of catalog2 at percentage $p_2.$ This is done for all pairwise combinations of samples ($6$ combinations if catalog1 $=$ catalog2, and $9$ combinations otherwise), resulting in a distribution of $p$-values of statistical significance.

We also compare computations for three samples of full data sets $\mathbf{(100\%)}$ such that the order of edges of
same lengths in the matrix reduction algorithm are shuffled. Such a shuffling will have no effect on
the persistence diagram, but the computed tight representatives might be different. These samples
are shown as $100\%$ shuffling in the figures. The birth and persistence thresholds for all data
sets are $\tau_u=22.5 \, h^{-1} \, {\rm Mpc}$ and $\epsilon=7.5 \, h^{-1} \, {\rm Mpc},$
respectively. 

\begin{figure}
\centering
\includegraphics[width=\textwidth]{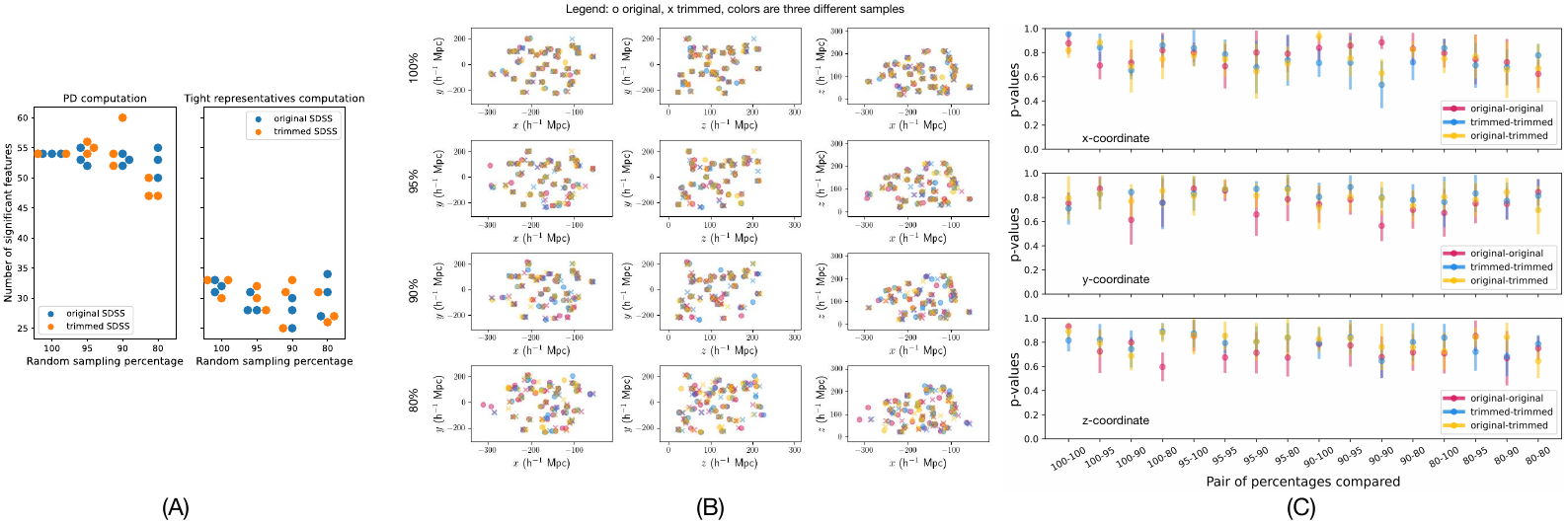} 

\caption{\label{fig:samethreshold} Comparing PH and tight representatives of different samples of
  original and trimmed data sets at different sampling percentages.  (A) Number of significant
  features in PD and number of tight representatives computed when birth and persistence thresholds
  were kept the same across all samples ($\tau_u = 22.5 \, h^{-1} \, {\rm Mpc}, \epsilon = 7.5 \,
  h^{-1} \, {\rm Mpc}$). (B) Centroids of computed voids. (C) Results of the Mann-Whitney U rank test
  for comparing distributions of $x, y $ and, $z$ coordinates of void centroids for all pairwise
  combinations show that the these distributions cannot be statistically distinguished ($p$-values
  $\gg 0.05$).}

\end{figure}

Figure \ref{fig:samethreshold}A shows the number of significant features from PD computation and the
number of computed tight representatives. The number of significant features from PD computation in
the full original and trimmed data sets are exactly the same, as expected. The number of computed
tight representatives varies between $30$ and $33$ for the different shuffled full data sets of
original and trimmed. The variance in the number of significant features and number of tight
representatives increases at percentages $90$ and $80.$

We next compare the centroids of the computed tight representatives. Figure
\ref{fig:samethreshold}(B) shows coordinates of the centroids for all cases---original is shown by o
marker, trimmed by x marker, and different colors show the different samples. We note that majority
of o and x match and the pattern is also similar as sampling percentage decreases (comparing top to
bottom). We quantify this by conducting Mann-Whitney U rank test for $x, y,$ and $z$ coordinates.
Figure \ref{fig:samethreshold}(C) shows that the $p$-values from this significance test are $\gg
0.05,$ indicating that these distributions cannot be statistically distinguished.

\section{Information for void centroids}

        Table~\ref{tab:multicol} shows centroids, $R_{\text{eff}},$ and sphericity of the tight representatives computed for PH voids in SDSS. The three different samples (Sample $0, 1,$ and $2$) correspond to tight representatives computed for three different bijective mappings of galaxies to integers. Such permutations will not affect the persistence diagrams but can result in differences in the tight representative computation due to the stochasticity of our algorithm. Nevertheless, Figure~\ref{fig:samethreshold}C shows that these differences are not significant---$p$-values for $100$-$100$ comparison for both original and trimmed are much greater than $0.05$ and very close to $1.$ The main text uses results of Sample $1.$

\begin{table}[ht]
\caption{SDSS void centroids of the computed tight representatives}
\begin{center}
\begin{tabular}{ccccc|ccccc|ccccc}
    \hline
    \multicolumn{5}{c}{Sample 0} & \multicolumn{5}{c}{Sample 1} &\multicolumn{5}{c}{Sample 2} \\
    \hline
    ra     &dec   &$z$    &$R_{\text{eff}}$ & $\psi$ & ra     &dec   &$z$    &$R_{\text{eff}}$ &
    $\psi$ & ra     &dec &$z$    &$R_{\text{eff}}$ & $\psi$\\
    \hline

236.62&26.3&0.101&43.62&0.78&235.8&25.63&0.1013&41.63&0.78&236.44&26.35&0.1008&44.04&0.74\\
172.34&33.83&0.0941&51.86&0.75&172.31&34.23&0.0941&54.84&0.73&174.59&34.82&0.0938&46.64&0.77\\
125.57&44.47&0.0872&44.05&0.8&211.55&43.61&0.0539&47.69&0.77&125.64&46.12&0.0885&42.87&0.8\\
196.07&21.55&0.0975&48.33&0.75&125.14&44.77&0.0879&41.86&0.78&155.34&21.14&0.0926&43.15&0.77\\
224.22&4.4&0.1048&33.8&0.8&217.43&9.46&0.0752&40.69&0.76&213.5&43.31&0.0522&46.3&0.74\\
210.25&43.62&0.0519&45.78&0.77&223.64&3.97&0.1051&34.89&0.78&217.26&9.45&0.0757&40.77&0.74\\
216.85&10.13&0.0754&39.14&0.74&193.08&22.35&0.0977&43.0&0.8&224.47&4.25&0.1035&33.52&0.84\\
154.99&22.23&0.0965&43.53&0.75&213.6&30.73&0.0953&40.3&0.81&193.14&21.55&0.0975&44.04&0.81\\
134.56&46.17&0.0622&36.01&0.8&156.97&21.87&0.0947&40.82&0.78&213.39&30.56&0.0972&41.9&0.8\\
212.36&29.64&0.0981&41.9&0.82&201.06&20.25&0.0565&34.75&0.8&134.12&45.49&0.062&35.97&0.8\\
149.28&12.51&0.0911&39.62&0.78&133.31&45.48&0.0608&35.39&0.79&157.98&19.34&0.0732&32.58&0.81\\
157.14&29.53&0.1061&33.64&0.85&148.39&13.01&0.0916&38.77&0.82&157.26&28.93&0.1057&33.84&0.83\\
199.41&19.34&0.0544&35.75&0.8&239.96&12.7&0.0823&34.68&0.83&199.66&19.4&0.0555&34.17&0.79\\
157.37&18.15&0.0732&35.9&0.81&142.68&17.49&0.0618&35.19&0.83&143.46&17.33&0.0607&33.89&0.82\\
155.57&30.59&0.0604&38.61&0.79&157.08&29.31&0.1058&33.3&0.86&233.75&13.08&0.0652&34.23&0.83\\
142.42&17.74&0.062&35.29&0.79&155.28&33.37&0.0628&38.15&0.77&179.27&28.27&0.0644&27.86&0.81\\
152.01&38.67&0.0992&28.13&0.86&152.84&38.95&0.0996&29.47&0.79&152.83&30.89&0.0607&36.6&0.81\\
126.78&42.1&0.0728&29.74&0.83&179.29&29.21&0.0634&27.57&0.84&169.64&38.13&0.0532&30.92&0.83\\
234.06&13.46&0.067&29.46&0.86&126.73&41.27&0.0725&30.39&0.83&152.83&38.78&0.0972&27.37&0.86\\
177.87&28.67&0.0642&27.26&0.83&233.83&10.2&0.0641&27.92&0.85&135.28&9.31&0.0999&30.76&0.8\\
194.05&14.33&0.1058&32.79&0.82&232.06&44.66&0.0828&31.85&0.81&195.01&60.23&0.0844&33.1&0.88\\
192.74&60.28&0.0826&33.17&0.86&192.45&60.6&0.0847&34.79&0.84&214.6&12.66&0.0501&30.53&0.85\\
189.11&35.92&0.0751&31.07&0.82&193.04&14.69&0.1061&33.78&0.82&193.07&14.78&0.1062&35.24&0.82\\
217.38&10.37&0.0466&30.04&0.83&211.92&11.65&0.052&28.05&0.85&231.51&44.58&0.0837&31.65&0.86\\
201.79&27.99&0.0811&26.85&0.84&134.48&10.15&0.0996&31.12&0.79&126.12&41.95&0.0735&28.08&0.87\\
230.87&45.3&0.0838&30.41&0.88&200.33&28.02&0.0822&27.7&0.87&189.06&37.16&0.0773&28.06&0.85\\
136.2&8.66&0.1004&28.79&0.82&227.21&25.87&0.0936&30.07&0.78&201.51&28.02&0.0812&26.9&0.81\\
132.45&31.73&0.0345&27.31&0.89&189.94&37.18&0.0779&26.84&0.87&227.76&24.92&0.096&28.03&0.81\\
151.46&36.82&0.0858&25.5&0.91&151.04&38.06&0.0841&27.28&0.89&150.8&37.11&0.0852&26.49&0.88\\
141.57&47.16&0.0992&18.82&0.94&132.26&30.77&0.0341&27.66&0.88&127.11&27.94&0.0613&26.51&0.81\\
205.06&24.72&0.1&22.88&0.9&206.11&25.47&0.0987&22.12&0.91&133.42&32.51&0.0353&26.87&0.92\\
-&-&-&-&-&142.38&46.01&0.0993&19.68&0.92&205.26&24.53&0.0991&25.44&0.91\\
-&-&-&-&-&-&-&-&-&-&141.95&45.46&0.0995&20.05&0.91\\
      \hline
\end{tabular}
\end{center}
\label{tab:multicol}
\end{table}

\section{Supplementary information}

\subsection{Computation cost}\label{sec:supp_computation_cost}

\begin{figure}
    \centering
    \includegraphics[width=0.75\linewidth]{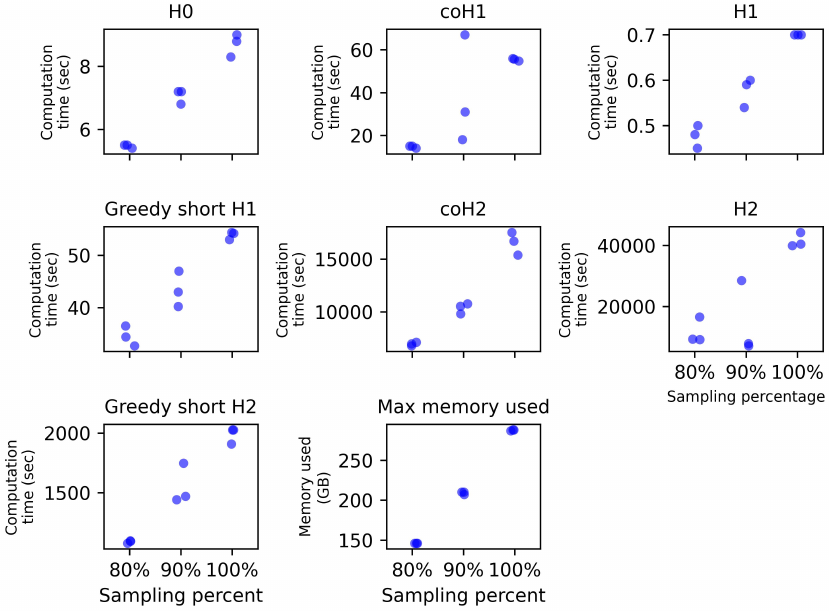}
    \caption{Computation times and memory for different under samples of the SDSS data set.}
    \label{fig:supp_computation_cost}
\end{figure}

Figure \ref{fig:supp_computation_cost} shows computation costs for the different steps of computing tight representatives (up to spatial scale of $30 \, h^{-1}$ Mpc and greedy shortening).

\bibliography{biblio}{}
\bibliographystyle{aasjournal}

\end{document}